\documentclass[12pt,preprint]{aastex}

\newcommand{\ltsimeq}{\raisebox{-0.6ex}{$\,\stackrel
        {\raisebox{-.2ex}{$\textstyle <$}}{\sim}\,$}}

\newcommand{\gtsimeq}{\raisebox{-0.6ex}{$\,\stackrel
        {\raisebox{-.2ex}{$\textstyle >$}}{\sim}\,$}}

\shortauthors{Gehrz et al.}

\shorttitle{``Neon Abundance in QU Vul''}

\begin{document}

\title{THE NEON ABUNDANCE IN THE EJECTA OF QU VUL
FROM LATE-EPOCH IR SPECTRA} 

\author{
ROBERT D. GEHRZ\altaffilmark{1},
CHARLES E. WOODWARD\altaffilmark{1}, 
L. ANDREW HELTON\altaffilmark{1}, 
ELISHA F. POLOMSKI\altaffilmark{1}, 
THOMAS L. HAYWARD\altaffilmark{2}, 
JAMES R. HOUCK\altaffilmark{3}, 
A. EVANS\altaffilmark{4},
JOACHIM KRAUTTER\altaffilmark{5}, 
STEVEN N. SHORE\altaffilmark{6}, 
SUMNER STARRFIELD\altaffilmark{7}, 
JAMES TRURAN\altaffilmark{8}, 
G.~J. SCHWARZ\altaffilmark{9}, 
R. MARK WAGNER\altaffilmark{10}
} 

\altaffiltext{1}{Department of Astronomy, School of Physics and Astronomy, 
116 Church Street S.E., University of Minnesota, Minneapolis, Minnesota 55455,
\ \it{gehrz@astro.umn.edu}}

\altaffiltext{2}{Gemini Observatory, Southern Operations Center, Casilla 603, 
La Serena, Chile} 

\altaffiltext{3}{Department of Astronomy, Cornell University, Ithaca, NY 
14853} 

\altaffiltext{4}{Astrophysics Group, Keele University, Keele, Staffordshire 
ST5, 5BG, UK} 

\altaffiltext{5}{Landessternwarte, K\"{o}nigstuhl, D--69117, Heidelberg, 
Germany} 

\altaffiltext{6}{Dipartimento di Fisica ``Enrico Fermi,'' Universita' di 
Pisa, largo Pontecorvo 3, Pisa 56127 Italy; INFN - Sezione di Pisa} 

\altaffiltext{7}{School of Earth and Space Exploration, Arizona State 
University, P.O. Box 871404, Tempe, AZ 85287} 

\altaffiltext{8}{Department of Astronomy and Astrophysics, University of 
Chicago, 5640 S. Ellis Avenue, Chicago, IL 60637 and Argonne National 
Laboratory, 9700 South Cass Road, Argonne, IL 60439}

\altaffiltext{9}{Department of Geology and Astronomy, West Chester 
University, 750 S. Church St., West Chester, PA 19383} 

\altaffiltext{10}{Large Binocular Telescope Observatory, University
of Arizona, 933 North Cherry Avenue, Tucson, AZ 85721}


\begin{abstract}

We present ground-based SpectroCam-10 mid-infrared, MMT optical, and \textit{Spitzer} Space
Telescope IRS  
mid-infrared spectra taken 7.62, 18.75, and 19.38 years respectively after the outburst of the old 
classical nova QU Vulpeculae (Nova Vul 1984\#2).  The spectra of the ejecta 
are dominated by forbidden line emission from neon and oxygen.   
Our analysis shows that neon was, at the first and last epochs respectively, 
more than 76 and 168 times overabundant by number with respect to hydrogen 
compared to the solar value.  These high lower limits to the  neon abundance 
confirm that QU Vul involved a thermonuclear runaway on an ONeMg white dwarf 
and approach the yields predicted by models of the nucleosynthesis in such 
events. 

\end{abstract} 

\keywords{circumstellar matter: classical novae, cataclysmic variables -- 
stars: individual (QU Vul, Nova Vul 1984\#2)}

\clearpage

\section{Introduction}
\label{sec:intro}

\cite{Gehrzet85} discovered a very strong [Ne II]12.8~\micron \  forbidden 
emission line in the classical nova QU Vulpeculae (Nova Vul 1984\#2) 140 days 
after its outburst. The presence of this strong neon line and the subsequent 
appearance of additional infrared metallic forbidden emission lines in the 
spectrum as QU Vul evolved \citep{Gehrzet86, Greenhouse88} confirmed that 
there are oxygen-neon-magnesium (ONeMg) white dwarfs (WDs) in classical nova 
binary systems whose eruptions could significantly enrich the local 
interstellar medium \citep{Gehrz88, Gehrz98, Gehrz99, Gehrz02, Gehrz08}. 
\cite{Gehrz98} argued that theoretical simulations of nova thermonuclear 
runaways (TNRs) provide evidence that ONeMg novae may, in fact, compete with 
supernovae as sources of some CNONeMgAl isotopes.  We have pursued a long 
term program to measure metal abundances in nova ejecta using 
infrared (IR) spectroscopy of forbidden emission lines to test this 
hypothesis. 

Measurements of the integrated fluxes of these electron-impact 
(collisionally) excited, forbidden lines in nova ejecta many years after 
outburst are particularly valuable for evaluating abundances. In this case, 
the density of the expanding ejecta will have fallen well below the level 
where the lines are quenched by collisions, and the line fluxes are entirely 
due to spontaneous emission. Since coefficients of spontaneous emission are 
fairly well known for many important IR lines, the absolute numbers of atoms 
involved in these transitions can be calculated with reasonable certainty given a known distance. 
Because the ejected mass of hydrogen is well known from early infrared and 
radio emission measures when most of the gas is ejected and when it is almost 
entirely ionized \citep{Gehrz99, Gehrz02}, the ratio of the mass of an 
element required to produce the observed IR forbidden emission lines to the 
ejected hydrogen mass gives a fairly well constrained lower limit to the 
abundance of that element. When emission from all the lines from all 
ionic species and all isotopes of an element can be observed and evaluated, 
the lower limit converges to the actual abundance. 

Several mid-IR neon forbidden emission lines in QU Vul are sufficiently 
strong that we have been able to observe them using SpectroCam-10 on 
the 5 m Hale Telescope at Mount Palomar and the Infrared Spectrograph (IRS) on the
\textit{Spitzer} Space Telescope at epochs of 
7.62 and 19.38 years after the eruption.  These measurements have enabled us 
to determine a fairly strong lower limit to the  neon abundance in this nova 
as reported below. 

\section{OBSERVATIONS AND REDUCTION}
\label{sec:obs}

\subsection{Ground-based Mid-IR}
\label{sec:gbir}

\begin{figure}
\plotone{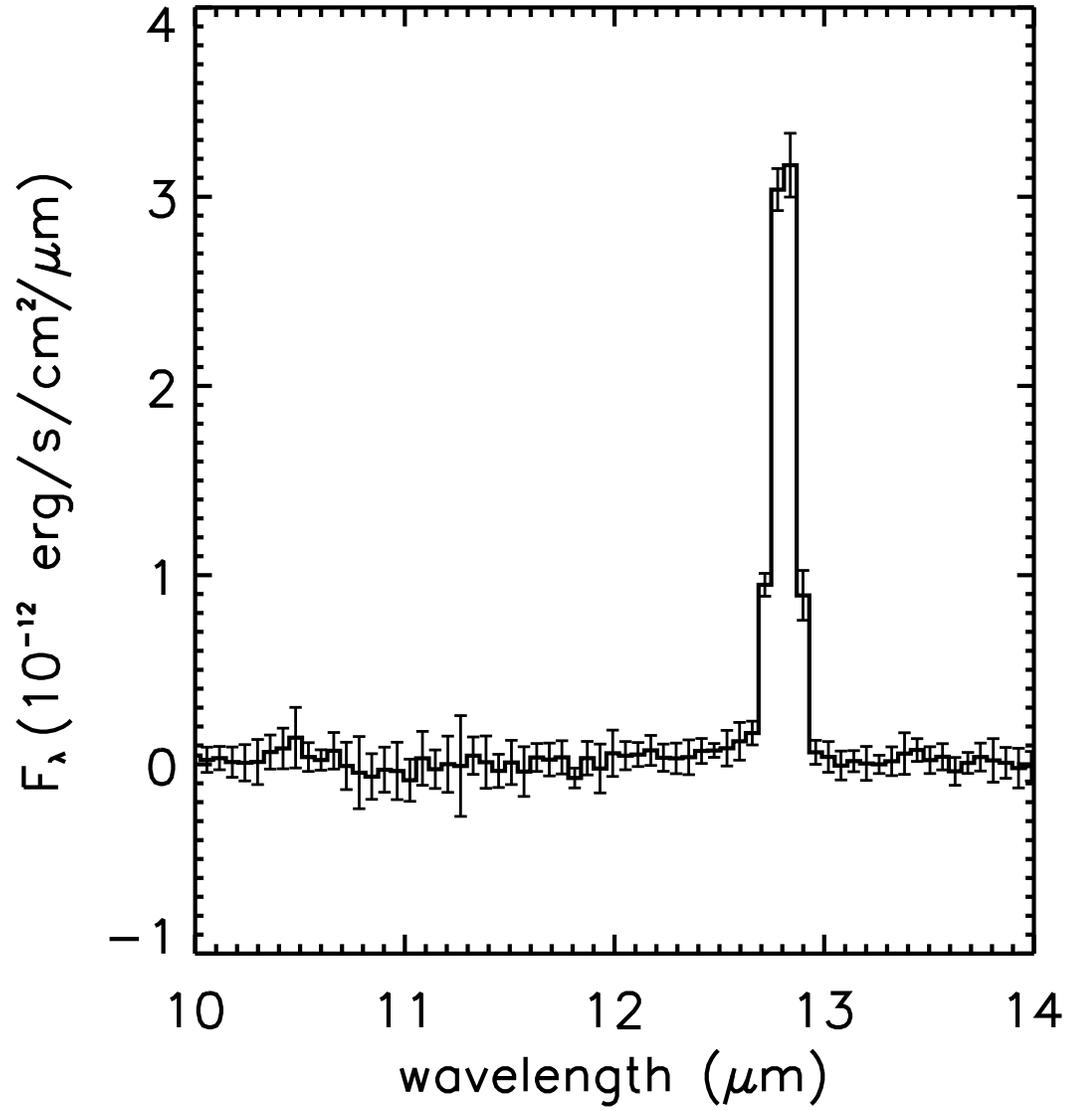}
\caption{SpectroCam-10 spectrum of QU Vul taken on 1992 August 09.3~UT.  
The [Ne II] line at 12.81~\micron \ has an integrated flux of $6.84 \times
10^{-12}$~erg~s$^{-1}$~cm$^{-2}$. 
\label{fig:midir_sc10}}
\end{figure}

We observed QU Vul with the 200 inch Hale Telescope at Mount Palomar
\footnote{Observations at
the Palomar Observatory were made as part of a continuing collaborative agreement between the
California Institute of Technology and Cornell University.}  on 1992 August 09.3~UT 
(day 2783.9, 7.62 years post-eruption) using the SpectroCam-10 (SC-10) mid-IR 
camera/spectrograph \citep{Hayward93} in its long-slit, low-resolution 
spectroscopy mode with a 2\arcsec \ wide slit and a resolving power of $R$ = 
$\lambda/\Delta\lambda \sim 150$ at 12.81~\micron.  We note that the entire extent of the
ejected shell  was contained within the slit (see Krautter et al. 2002).  A single grating setting 
provided a 10.6 to 13.2~\micron \ spectrum that included the [Ne 
II]12.81~\micron \ line.  The  total on-source integration times were 160~s on 
QU Vul and 20~s on the standard star $\alpha$~Lyr. The data were reduced by 
extracting  1-dimensional spectra from the 2-dimensional spectral images, 
dividing the  nova by the standard, then multiplying by the 
absolute spectral energy distribution of $\alpha$~Lyr \citep{Cohen92}.  The 
calibrated result, with 2-pixel binning for clarity, is shown in 
Figure~\ref{fig:midir_sc10}.  In Table~\ref{tab:neon_ab_tab}, we give the 
total flux measured in the line. 

\subsection{Ground-based Optical}
\label{sec:gbopt}

\begin{figure}
\plotone{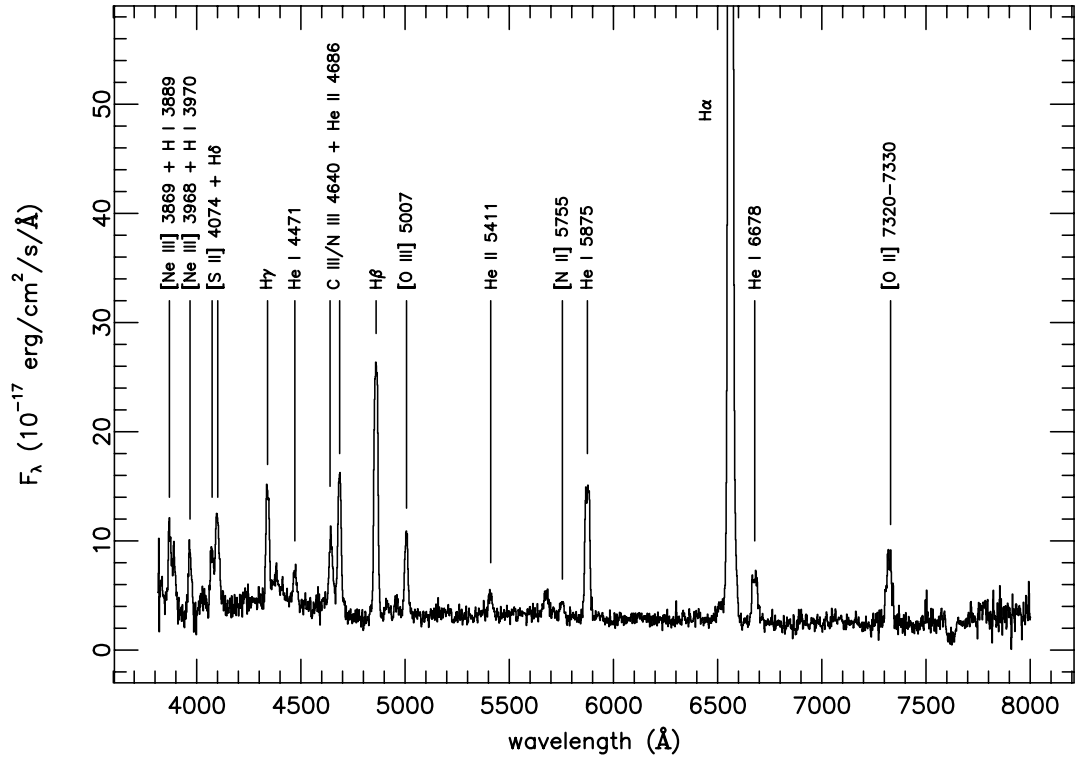}
\caption{MMT Blue Channel optical spectrum of QU Vul taken on 2003 
September 13.31~UT. Prominent recombination lines of H and He are evident.  The [Ne~III]
lines near 3869 and 3968~\AA \ are weak and blended with 
adjacent hydrogen lines at the spectral resolution of the instrument. 
\label{fig:mmt_blu}}
\end{figure}

An optical spectrum of QU~Vul was obtained with the 6.5~m Multiple Mirror Telescope  and
Blue  
Channel Spectrograph \citep{SWF89} on 2003 September 13.31 UT  (day 6836;
18.72~years post-eruption) in  
generally photometric conditions and $1\arcsec$ FWHM seeing.  Three  
600~s integrations of the nova were obtained through a $1\arcsec  
\times 180\arcsec$ entrance slit oriented along the parallatic angle  
to avoid loss of light at the slit from differential atmospheric  
refraction and to improve the relative photometric accuracy. A 300  
lines/mm grating centered at 6400 \AA\ was utilized which covered the  
entire optical region on the detector; however, a long--pass order  
separation filter blocked all light shortward of $\sim$3800 \AA\ and  
yielded calibrated spectra covering the region 3800 -- 8000 \AA\ at a  
resolution of about 7 \AA\  FWHM.  The spectrophotometric standard  
star BD+28 4211 was observed at a similar airmass as QU Vul using a  
wider $5\arcsec \times 180\arcsec$ entrance slit to enable accurate  
flux calibration.  The spectra of HeNeAr and a quartz--halogen lamps  
were obtained to facilitate accurate wavelength calibration and  
flatfielding respectively.  The spectra were reduced using standard  
IRAF packages\footnote{IRAF is distributed by the National Optical Astronomy Observatories,
which are operated by the Association of Universities for Research in Astronomy, Inc., 
under cooperative agreement with the National Science Foundation.} and spectral
extraction techniques.    The optical spectrum is
presented in Fig.~\ref{fig:mmt_blu}. 

\subsection{\textit{Spitzer} \  IRS}
\label{sec:spz_irs}

We observed QU Vul on 2004 May 11.7~UT (day 7077.6, 19.38 years 
post-eruption) with the IRS \citep{Houck04} on the {\it{Spitzer}} Space 
Telescope \citep[\textit{Spitzer},][]{Werner04,Gehrzet07} as  part of the 
Gehrz Guaranteed Time Observing Program (GGTOP), program identification 
number (PID) 124.  The observations were conducted using the short wavelength 
(5.2 -- 14.5~\micron)  low resolution module (SL), the  short wavelength 
(9.9 - 19.6~\micron)  high resolution module (SH), and the long  wavelength 
(18.7 -- 37.2~\micron) high resolution module (SH).   The resolving power for SL 
is $R$ = $\lambda/\Delta\lambda \sim 60-120$, while R $\sim$ 600 for SH and 
LH. Emission lines observed with the latter modules are marginally resolved.  
All observations utilized visual PCRS peakup on nearby isolated stars to 
ensure proper placement of the target in the narrow IRS slits.  The 
spectroscopic astronomical observation request (AOR) key
\dataset[ADS/Sa.Spitzer#0005044992]{0005044992}, 
consisted of 5 cycles of 6 second ramps in SL (30 seconds on-source), and 6 
cycles of 6 second ramps in SH and LH (36 seconds on-source each). 

\begin{figure}
\plotone{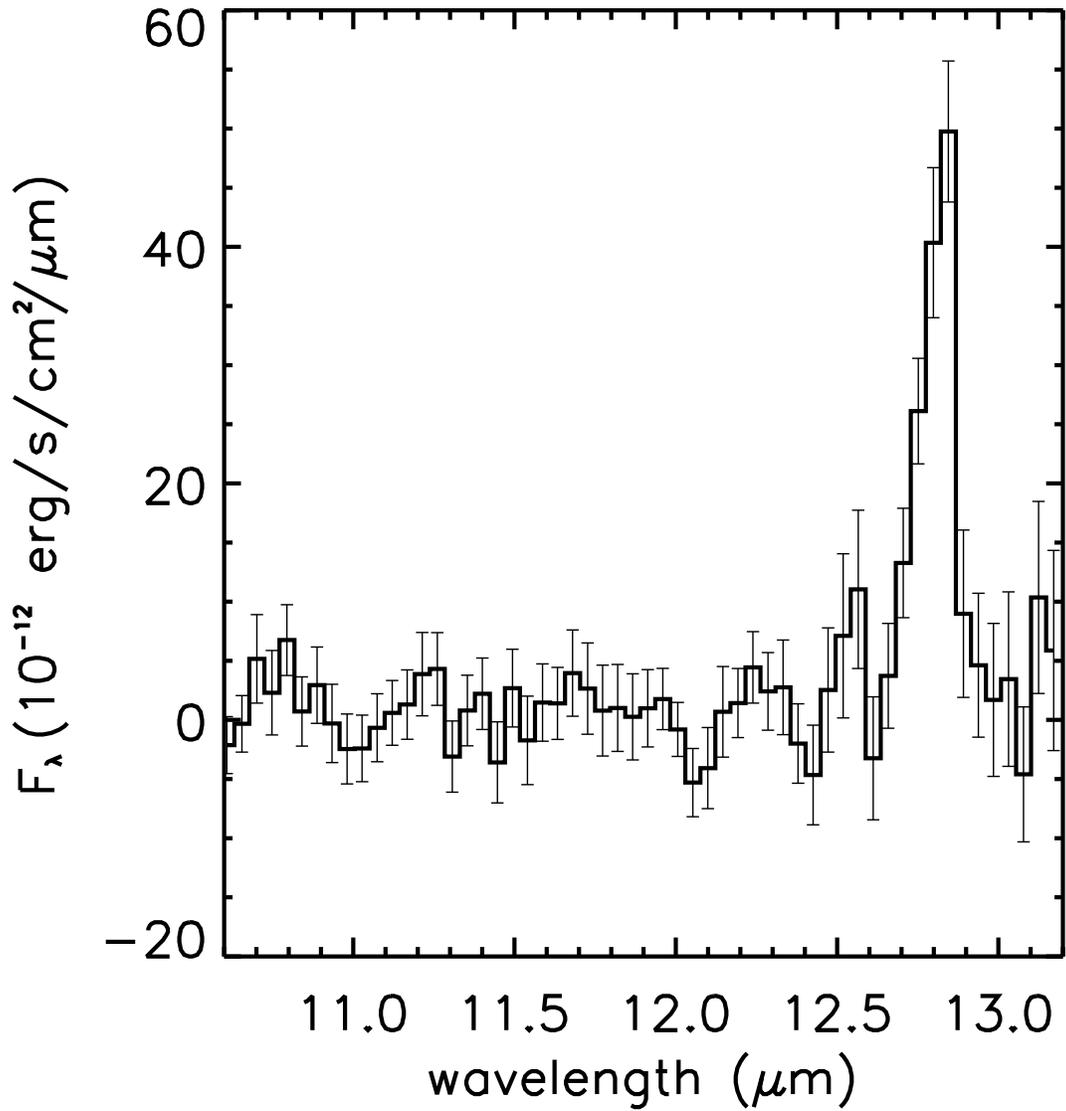}
\caption{\textit{Spitzer} background-subtracted Short Low (SL) spectrum of QU 
Vul taken on 2004 May 11.7~UT.  The [Ne II] line at 
12.81~\micron \ has an integrated line flux of $4.96 \times 10^{-13}$ erg 
s$^{-1}$ cm$^{-2}$.
\label{fig:irs_sl}}
\end{figure}

\begin{figure}
\plotone{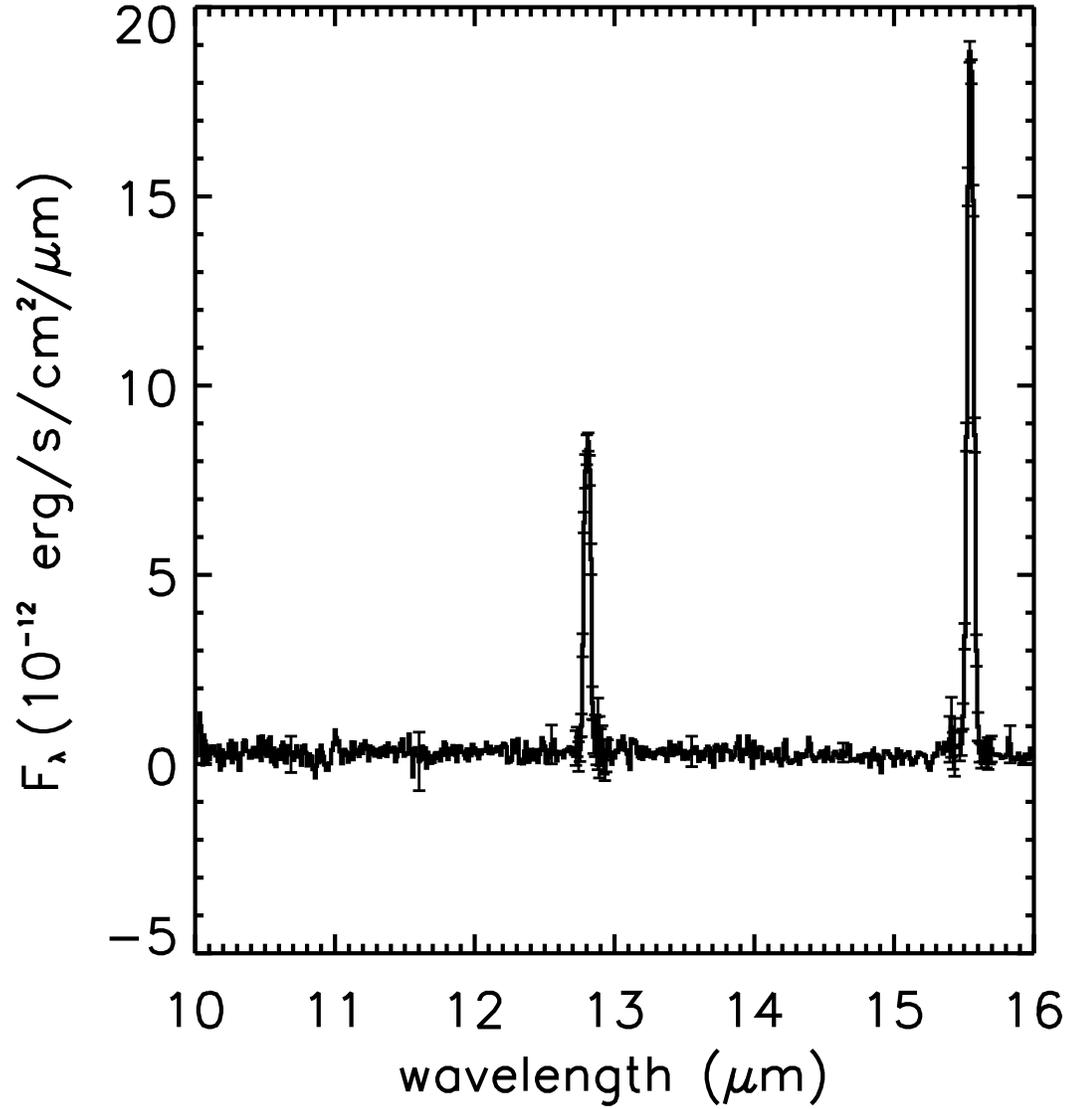}
\caption{ \textit{Spitzer} Short High (SH) spectrum of QU Vul taken on 2004 May 
11.7~UT.  No sky subtraction was performed.  The [Ne II] line at 12.81~\micron 
\ has an integrated line flux of $4.34 \times 
10^{-13}$~erg~s$^{-1}$~cm$^{-2}$.  Also evident is the [Ne III] line at 
15.56~\micron \ with an integrated line flux of $1.09 \times 
10^{-12}$~erg~s$^{-1}$~cm$^{-2}$.  For clarity, a representative subset of 
continuum errors is plotted. 
\label{fig:irs_sh}}
\end{figure}

\begin{figure}
\plotone{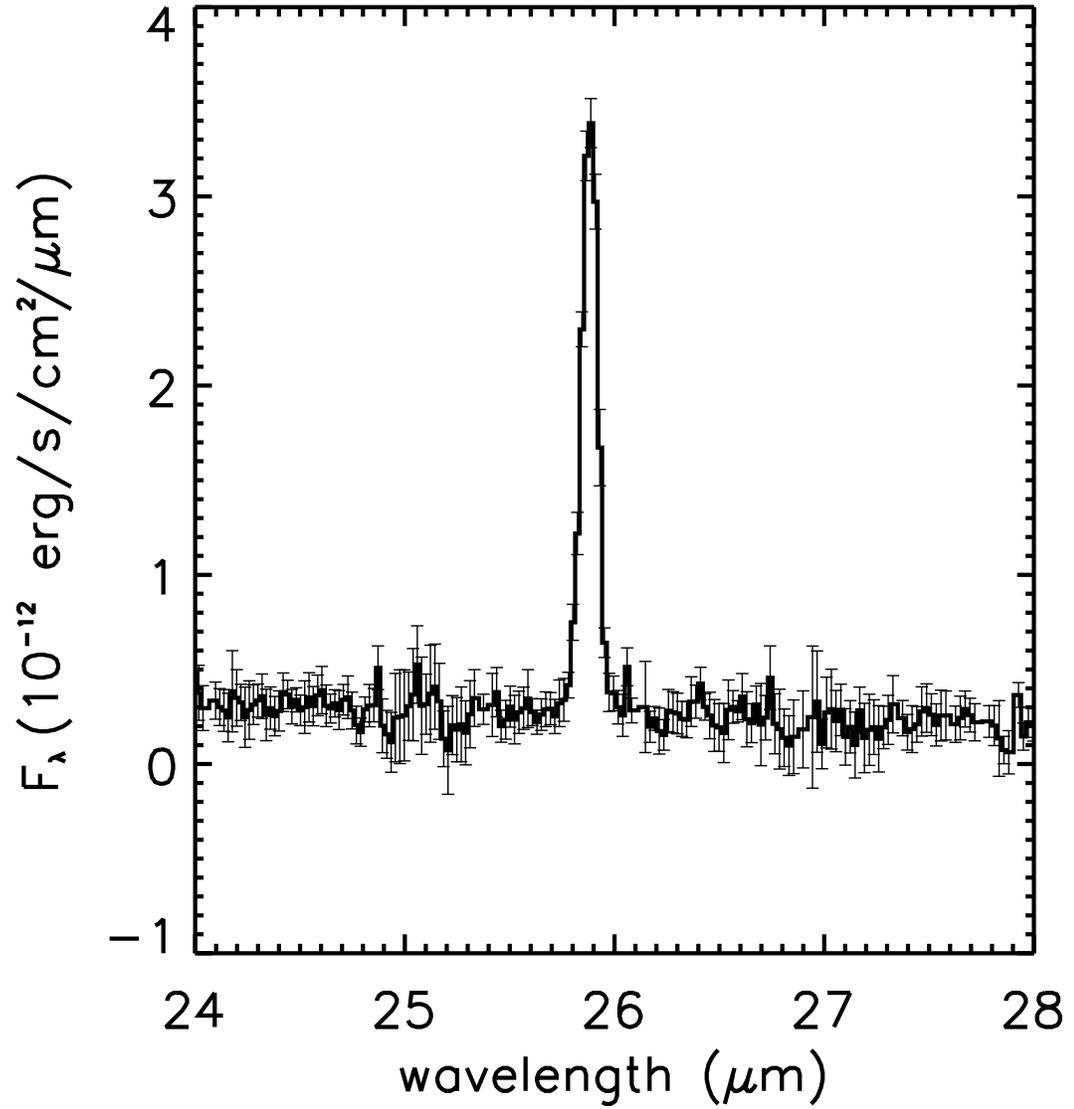}
\caption{ \textit{Spitzer} Long High 1 (LH1) spectrum of QU Vul taken on 2004 
May 11.7~UT.  No sky subtraction was performed.  The
[O IV] line at 25.91~\micron \ has an integrated line flux of $3.04 \times 
10^{-13}$~erg~s$^{-1}$~cm$^{-2}$.  The absolute flux calibration of the flux near the line is
uncertain by less than 10\%.
\label{fig:irs_lh}}
\end{figure}

IRS Basic Calibrated Data (BCD) products were calibrated and processed with 
the {\it{Spitzer}} Science Center (SSC) IRS pipeline version 13.2. Details  
of the calibration and raw data processing are specified in the IRS Pipeline 
Description Document, 
v1.0.\footnote{http://ssc.spitzer.caltech.edu/irs/dh/PDD.pdf}  Bad pixels 
were interpolated over in individual BCDs using bad pixel masks provided by 
the SSC.  Multiple data collection events were obtained at two different 
positions on the slit using \textit{Spitzer's} nod functionality.  Sky 
subtraction was only possible for the SL observations as no dedicated sky 
observations were performed for the SH and the LH mode observations. Sky 
subtraction was performed by differencing the two dimensional SL BCDs to 
remove the background flux contribution.  Spectra were then extracted from 
the background corrected SL data and the SH and LH BCDs with SPICE (version 
1.3-beta1) using the default point source extraction widths. The extracted 
spectra were then combined using a weighted linear mean into a single output 
data file.  The high resolution IRS data were not defringed.   At the 
time of reduction, the errors generated by the SSC pipeline were not reliable 
and therefore errors were estimated from the standard deviation of the flux 
at each wavelength bin. The continuum was not detected over much of the 
spectral range of our data (absolute flux calibration of the IRS modules is \ltsimeq 10\%) and no
emission was detected in the second order 
SL spectra.  A good fit to the spectral lines was obtained using a non-linear 
least squares Gaussian  routine \citep[the Marquardt method,][]{Bev92} that 
fit the line  center, line amplitude, line width, continuum amplitude and the 
slope of the continuum.  The spectra are shown in 
Figs.~\ref{fig:irs_sl} through \ref{fig:irs_lh}.  
     
Given that the [Ne II]12.81~\micron \ line is accessible from the ground, we have investigated
the possibility of continuing to monitor it in QU Vul using today's very large ground-based
telescopes.  During 
the interval between our SpectroCam-10 measurement of QU Vul (7.62 years 
after outburst) and our \textit{Spitzer} measurement (19.32 years after 
outburst), we attempted unsuccessfully several times to detect this line in 
QU Vul using the long wavelength spectrometer LWS \citep{Jon93} on the Keck I 
10 m telescope.  More 
advanced mid-IR spectrometers are now available.  One example is the VLT Imager 
and Spectrometer for mid-IR (VISIR) at ESO Paranal.   We have used the 
VISIR Exposure Time Estimator (http://www.eso.org/observing/etc/) to 
calculate that VISIR on the ESO Paranal UT3 8.2 m telescope would 
require about 15 hours of integration time to achieve the same 
signal-to-noise ratio (20/1) on this line as we reached with \textit{Spitzer} IRS 
SL in 30 seconds.  We conclude that further ground-based monitoring of the 
[Ne~II]12.81~\micron \ line in QU Vul is untenable. 

\section{DETERMINATION OF ABUNDANCES}
\label{sec:abundances}

Lower limits to the abundance of an element in the ejecta of a nova can be 
determined by dividing the number of atoms required to produce the lines of 
the element that are observed by the total number of hydrogen atoms in the 
ejecta.  The lower limit so determined converges to the actual abundance when 
all lines of all ions and isotopes can be evaluated.  At very late epochs, 
hydrogen emission lines and the free-free continuum are no longer observed in the IR, 
so that the number of hydrogen atoms must be determined from the mass known 
to have been ejected in the eruption.  We note that the majority of the mass 
is expelled within the first few hours of the outburst and very little 
thereafter, so that the number of hydrogen atoms is well constrained by early 
IR measurements and radio continuum observations \citep{Gehrz88, Gehrz98, Gehrz99,
Gehrz02, Gehrz08}. 

The neon and oxygen emission lines observed in the spectra of QU Vul are 
forbidden transitions excited by electron impact. The total number of ions 
required to produce the observed line strengths, assuming a uniform 
temperature distribution and density along the line of sight in a homogeneous 
medium, can be determined following detailed balancing arguments described by 
\citet{Osterbrock89}.  The [Ne II], [Ne VI], and [O IV] lines can be 
treated as arising from two-level atoms.  [Ne III] has a $^{3}$P ground term giving rise 
to emission lines at 15.56 and 36.02~\micron.  Our spectra do not extend to 
long enough wavelengths to measure the strength of the 36.02~\micron \  line. 
In this case, we must treat the 15.56~\micron \  line as coming from a 
two-level atom, an approximation that gives us a lower limit to the number of 
Ne III atoms in the ejecta.  

The level populations (total number of ions), $N_{i}$ and $N_{j}$, in the 
lower $(i)$ and upper $(j)$ energy states can be computed from the line 
luminosity, $L_{c} = 4 \pi D^{2} F$, where $D$(cm) is the distance to 
the nova and $F$ (erg~s$^{-1}$~cm$^{-2}$) is the integrated apparent intensity 
of the line. For QU Vul, we take D to be 3.14 kpc \citep{Krautter02}.  We regard this distance
estimate as being superior to others given in the literature because the shell was well resolved
by the HST NICMOS imager allowing a well constrained expansion parallax to be derived. The 
total number of ions in the state $N_{i}$ is 

\begin{equation}
N_{i} = \frac{L_{c}}{n_{e}h\nu _{ji}}\frac{(1 + [n_{e}/n_{crit}])}{q_{ij}} 
\label{N-I}
\end{equation}

\noindent where $n_{e}$~(cm$^{-3}$) is the electron number density, 
$n_{crit}$ is the critical density of the transition, $h$ is the Planck 
constant, and $\nu_{ji}$~(Hz) is the frequency of the transition. The 
parameter $q_{ij}$~(cm$^{3}$~s$^{-1}$) is the excitation rate coefficient, 
which is a function of temperature, and is expressed as 

\begin{equation}
q_{ij} = \frac{8.63 \times 10^{-6}}{\omega_{i} T_{e}^{1/2}} \Upsilon_{ij} 
\exp\left[\frac{- h\nu_{ij}}{kT_e}\right] \label{q-ex}
\end{equation}

\noindent where $\omega_{i}$ is the statistical weight of the lower state, 
$k$ is the Boltzmann constant, T$_{e}$~(K) is the electron temperature,  and 
$\Upsilon_{ij}$ is the dimensionless thermally-averaged effective collision 
strength \citep{Hummer93}. Values of $\Upsilon_{ij}$, which is temperature dependent, were
taken from 
\cite{PradPeng95}, \cite{SaraphTully94}, \cite{McLaugBell00}, and 
\cite{BlumPradhan92} for [Ne VI] $(^{2}P_{3/2} \rightarrow ^{2}P_{1/2})$, 
[Ne II] $(^{2}P_{1/2} \rightarrow ^{2}P_{3/2})$,  
[Ne III] $(^{3}P_{1}\rightarrow  ^{3}P_{2})$, and 
[O IV] $(^{2}P_{3/2} \rightarrow  ^{2}P_{1/2})$ respectively. 

The critical density, $n_{crit}$~(cm$^{-3}$), for a given transition is 
computed from the ratio of $A_{ij}$ to $q_{ji}$  \citep{Osterbrock89} where 
$A_{ij}$~(s$^{-1}$) is the radiative transition probability and 
$q_{ji}$~(cm$^{3}$~s$^{-1}$) is the temperature-dependent, de-excitation rate 
coefficient defined as 

\begin{equation}
q_{ji} = \frac{8.63 \times 10^{-6}}{\omega_{j} T_{e}^{1/2}} \Upsilon_{ij} 
\label{q-dex}
\end{equation}

\noindent where $\omega_{j}$ = (2J + 1) is the statistical weight of the 
upper state and J is the total angular momentum quantum number of the upper 
state. 

When $n_{e} << n_{crit}$, collisional de-excitation of the ion in state ($j$) 
is not significant compared with radiative decay.  As can be seen from Table 
1 (columns 8 and 9), this was the case for the ejecta of QU Vul when 
the {\it{Spitzer}} observations were made in 2004. $A_{ji}$ values obtained 
from the NIST database\footnote{http://www.physics.nist.gov} are $2.02 \times 
10^{-2}$~s$^{-1} (\pm 10\%)$  for [Ne~VI], $8.59 \times 10^{-3}$~s$^{-1} \ 
(\pm 10\%)$  for [Ne~II],  and $5.19 \times 10^{-4}$~s$^{-1} \ (\pm 3\%)$  
for [O~IV]. For [Ne~III], we adopt $A_{ji}$ = $5.84 \times 10^{-3}$~s$^{-1} \ 
(\pm 10\%)$ from \citet{Kram06}. 

The total number of ions in state $N_{j}$ is expressed as

\begin{equation}
N_{j} = \frac{L_{c}}{A_{ji} h\nu _{ji}}. \label{eqn:N-j}
\end{equation}

The electron number density, $n_{e}$, of the ejecta at the epoch commensurate 
with the given neon line observation was inferred by adopting a 
free-expansion model for the ejected shell.  Assuming spherical symmetry and 
a uniform shell with a filling factor of unity, 

\begin{equation}
n_{e}=\frac{3}{4\pi }\frac{M_{ej}}{R_{s}^{3}}\frac{1}{m_p}=\frac{3}{4\pi
}\frac{M_{ej}}{(V_{o}\ \Delta t)^{3}}\frac{1}{m_p}\label{eqn:erho}
\end{equation}

\noindent where $M_{ej} = 3.6 \times 10^{-4}~M_{\odot}$ \citep{Tay88} is the 
mass of the ejected shell of radius $R_{s}$~(cm), $V_{o}$ is the outflow 
velocity in cm~s$^{-1}$ taken to be 1190~km~s$^{-1}$ \citep{Krautter02, 
Rosino92}, $\Delta t$ is the time since maximum light (day 0 = 1984 December 
25.1 UT =  JD 2,446,059.6) in seconds, and $m_{p}$ is the  mass of the 
proton. The slow velocity chosen here is consistent with the line widths, 
which are marginally resolved by IRS SH and SC-10. We emphasize that this is 
the mean electron density inferred from the unity filling factor and that 
there is some evidence for structure in the ejecta from the asymmetry in the 
[Ne~II] and [O~IV] lines. 

Thus, a lower limit to the abundance of neon in QU Vul with respect to the 
Solar value given by the strength of a given neon forbidden line alone is: 

\begin{equation}
\frac{\left[ \frac{Ne}{H} \right]_{QU Vul}}{\left[ \frac{Ne}{H} \right]_{\odot}}\geqslant \left[
\frac{\frac{N_{i}+N_{j}}{N_{H}}}{7.413\times 10^{-5}} \right] .\label{eqn:nabs_sol}
\end{equation}

\noindent where the Solar abundance of neon by number 
$\left[Ne/H\right]_{\odot} = 7.413 \times 10^{-5}$ \citep{Lodders03}, and the 
total number of hydrogen atoms in the nova shell is taken as ($M_{ej}/m_p) = 
4.3 \times 10^{53}$. On dates when more than one neon forbidden line was 
observed, the individual abundances derived from eqn.~\ref{eqn:nabs_sol} are 
summed to provide an estimate of the total neon abundance. We stress that the 
total mass of neon in the ejecta of QU Vul, derived by the method described 
above, is a lower limit.  The oxygen abundance is calculated in a similar fashion, where the solar
abundance of oxygen is $\left[O/H\right]_{\odot} =  4.898 \times 10^{-4}$ \citep{Lodders03}. 


The observed neon line flux densities, log($T_{e}$), \ $n_{e}$\ , level populations, and derived
neon abundances are summarized in Table~\ref{tab:neon_ab_tab}. An initial value of
log($T_{e}$) = 4.0  was adopted for the abundance calculation for the first few hundred days.
Later, a value of log($T_{e}$) = 3.0  was adopted based on the continuum-subtracted hydrogen
Brackett-$\gamma$ to Paschen-$\alpha$ emission line ratio derived from the Hubble Space
Telescope (\textit{HST}) narrow-band observations of the ejecta obtained by \cite{Krautter02}
on day 2783.9 (= JD 2,448,843.5). Assuming Case-B conditions and the hydrogen recombination
line emissivities tabulated by \cite{StoreyHummer95}, the \textit{HST} [$HI(7-4)/HI(4-3)] =
7.27 \pm 0.46 \times 10^{-2}$ ratio suggests that log($T_{e}$) = 3.0 for a $n_{e} = 1.0 \times
10^{3}$~cm$^{-3}$. The latter $n_{e}$ is of the order expected from eqn~\ref{eqn:erho}. 

Examination of the line profiles in the optical spectrum (Fig.~\ref{fig:mmt_blu})  that we
obtained $\approx 242$~days prior to the \textit{Spitzer} observations lead us to conclude that
the Balmer, He I, and He II lines probably come from both the low temperature ejecta and from
the hot accretion disk in the binary system.  Given the resolution of our spectrum, it is
impossible to de-convolve the different components, and an analysis of the hydrogen line ratios
yields an upper limit to the electron density in the ejecta.  Dereddened \citep[assuming E(B-V) = 
0.6,][]{andrea94, saizar92} flux ratios derived from Gaussian fits to the prominent H
recombination lines in the day 6836 (= JD 2,452,895.8) spectrum yield $(H\beta/H\alpha) \simeq
0.29 \pm 0.05$ and $(H\gamma/H\alpha) \simeq 0.14 \pm 0.03$. Assuming Case-B conditions
and the hydrogen recombination line emissivities tabulated by \cite{StoreyHummer95}, the 
optical line ratios are consistent with log($T_{e}$) = 3.0 and $n_{e} \sim
200-1,000$~cm$^{-3}$.  Again, this range of values for  $n_{e}$ is consistent with the the value
we derive by applying eqn~\ref{eqn:erho}.

For comparison, we have re-calculated the neon abundances that can be 
inferred from the data taken on 1985 May 15.4 UT \citep{Gehrzet85}, 1985 
August 23.4 UT \citep{Gehrzet86}, and 1986 July 30.0 UT and August 11.0 UT 
\citep{Greenhouse88} based upon a more recent understanding of the expansion 
velocity and density development of the ejecta as described above, as well as 
the new values for the excitation and emission constants associated with the 
observed neon transitions.   As is evident from the data in Table~\ref{tab:neon_ab_tab}, the early
epochs clearly fail the criterion that  $n_{e} << n_{crit}$, suggesting that the dramatic difference
in abundance estimates for the early epoch relative to the later epochs likely is due to strong
collisional damping of the emission lines.  Also, it is not obvious that the relative locations of
line formation within the expanding ejecta have remained the same over time.

The large neon overabundances (at least 76 times solar on day 2783.9 
and 168 times solar on day 7077.6) derived from our most recent 
observations (Table 1, column 13) are even larger than the value of 
21.7 $\pm$ 1.7 reported by Schwarz (2002; hereafter S02) based on detailed 
photoionization models constrained by multi-wavelength line fluxes 
obtained much earlier in the development of the ejecta.  Closer 
examination of the assumptions made in the current analysis reveals that 
the discrepancies between the current analysis and that made by S02 can 
be easily resolved.  The best fit CLOUDY model parameters reported by 
S02 suggest a distance estimate of 2.4 kpc compared to our adopted value 
of 3.14 kpc.  From equations (1),  (4), and 
the the relationship between flux and luminosity, the abundance estimates scale as $D^{2}$.  If we 
adopt instead a distance of 2.4 kpc, our calculated abundances drop by a 
factor of $\approx$ 1.7.  Further, in 
our analysis, we have made the elementary assumption that the ejecta 
uniformly fills a spherical volume with $R = V_{0}\ \Delta t$.  If instead we assume 
a spherically symmetric shell with a total volume half that of the 
sphere approximation, the electron number density would go up by a 
factor of two consequently reducing our abundance estimates by a 
corresponding factor of two, bringing them into close agreement with 
estimates made by S02.  On the other hand, we note that S02 used a solar neon abundance 1.59
higher than the more current value given by Lodders (2003).  Correcting S02's values upward by
this factor brings them in line with our results given the distance and the volume estimates
discussed above.  We also stress that the abundance values for 
neon given by us are inferred from the total mass of hydrogen in the 
ejecta as determined from radio continuum measurements (Taylor et al. 
1988) and not from hydrogen recombination line strengths.

Other estimates of the neon abundance by number in QU Vul using 
ionization correction factor (ICF) techniques and CLOUDY modeling have 
ranged from 60 - 254 times solar.  The weaknesses in the ICF 
method and other CLOUDY models are discussed in detail by S02.  The 
reader should be aware that Livio \& Truran (1994) have pointed out that 
the uncertainties in the various methods of determining abundances in 
novae can lead to discrepancies of factors of  $\approx$ 2-4 under some 
circumstances.  Our results (Table 1), taken in context along side rigorous photoionization
modelling, suggest a significant overabundance of neon in the ejecta of QU Vul.

The neon that we have detected in our spectra is most likely $^{20}$Ne dredged-up from the
WD outer layers at some time during the thermonuclear runaway.  In addition, a fraction could be 
$^{22}$Ne which is produced from the decay of $^{22}$Na ($\tau_{1/2}$ = 2.6 yr) and which
is predicted to be produced by thermonuclear runaways on ONeMg WDs \citep{ss92}. 
Moreover, the formation and ejection of  low levels of $^{22}$Na, decaying into $^{22}$Ne,
can account for the production of some of the $^{22}$Ne $(Neon-E)$ anomalies found in
meteorites \citep{Black72}.

\section{CONCLUSIONS}
\label{sec:conl}

Given the data summarized in Table~\ref{tab:neon_ab_tab}, we conclude that 
neon in this nova was at least 76 and 168 times overabundant with respect to 
the solar value at epochs of 7.62 and 19.38 years past outburst.  While this 
indicates that individual novae such as QU Vul might be important local 
sources for enriching the interstellar medium with  neon, they likely 
contribute no more than a few percent of the total Galactic abundance of 
neon.  However, a fraction of the ejected neon could be $^{22}$Ne produced by the 
decay of $^{22}$Na which may be produced during the outburst and could contribute
to the $^{22}$Ne $(Neon-E)$ anomalies found in meteorites.  However, the 
most significant  implication of the high level of neon enrichment observed 
in QU Vul - since its source must certainly be dredge-up of  matter 
from the underlying degenerate core - is that it provides unambiguous 
evidence for the occurrence of ONeMg white dwarfs among cataclysmic variable 
systems.  The level of enrichment seen here is consistent with the high values 
predicted by the theoretical models of TNRs on ONeMg WDs presented by \citet{Ili02}. 

\acknowledgements

We thank J.  Jos\'e  and K. Vanlandingham for illuminating discussions.  
This work is based in part on observations made with the {\it{Spitzer}} Space  
Telescope, which is operated by the Jet Propulsion Laboratory, California 
Institute of Technology, under NASA contract 1407.  RDG, CEW, LAH, and EFP 
are supported in part by NASA through contracts 1256406, 1215746, and 1267992 
issued by JPL/Caltech to the University of  Minnesota.  SS acknowledges 
partial support to ASU from the NSF and NASA. JWT is supported in part by the 
NSF under Grant PHY 002-16783 for the Physics Frontier Center, Joint 
Institute for Nuclear Astrophysics and by the DOE, Office of Nuclear Physics 
under Grant W-31-1009-ENG-38. 

{\it Facilities:} 

\facility{Spitzer (IRS)}, 

\facility{200 inch Hale Telescope  (SpectroCam-10)}

\facility{MMT (Blue channel spectrograph)}

\clearpage


\clearpage

\begin{deluxetable}{llclcccccccc}
\tabletypesize{\scriptsize}
\setlength{\tabcolsep}{0.025in}

\tablewidth{0pt}
\tablecaption{Physical Parameters for Neon Emission Lines Observed in  
QU Vulpeculae \label{tab:neon_ab_tab}}

\tablehead{

\colhead{(1)} &

\colhead{(2)} &

\colhead{(3)}  &

\colhead{(4)} &

\colhead{(5)} &

\colhead{(6)} &

\colhead{(7)} &

\colhead{(8)} &

\colhead{(9)} &

\colhead{(10)} &

\colhead{(11)} &

\colhead{(12)\tablenotemark{a}}\\

\colhead{ } &

\colhead{ } &

\colhead{ } &

\colhead{Emission} &

\colhead{Integrated} &

\colhead{ } &

\colhead{ } &

\colhead{ } &

\colhead{ } &

\colhead{ } &

\colhead{ } &

\colhead{[$\frac{Z}{H}$]$_{QU Vul}$}\\

\colhead{ } &

\colhead{Date} &

\colhead{ } &

\colhead{Line}&

\colhead{Flux} &

\colhead{log~T$_{e}$} &

\colhead{ } &

\colhead{$n_{e}$} &

\colhead{$n_{crit}$} &

\colhead{$N_{i}$} &

\colhead{$N_{j}$}&

\colhead{ --------------- }\\

\colhead{OBS} &

\colhead{(UT)} &

\colhead{Day} &

\colhead{(\micron)} &

\colhead{($10^{-12}$ \ erg~s$^{-1}$~cm$^{-2}$)} &

\colhead{(K)} &

\colhead{$\Upsilon_{ij}$} &

\colhead{($10^{5}$ \ cm$^{-3}$)} &

\colhead{($10^{5}$ \ cm$^{-3}$)} &

\colhead{($10^{50}$)}&

\colhead{($10^{49}$)}&

\colhead{[$\frac{Z}{H}$]$_{\odot}$}\\

 }

\startdata

WIRO \ \ &1985 May 15.4\ \ \ &141.3\ \ \

&[Ne II] 12.81 &220$\pm$20 &4.0

&0.283 &332 &7.03 &4.5 &19.5&$\gtsimeq 20$\\

WIRO \ \ &1985 Aug 23.4\ \ \ &241.3\ \ \

&[Ne II] 12.81 &97$\pm$10 &4.0

&0.283 &67 &7.03 &2.1 &8.6 &$\gtsimeq 9.4$\\

IRTF \ \ &1986 Jul 30.0\ \ \ &581.9\ \ \

&[Ne II] 12.81 &71.9$\pm$3.4 &4.0

&0.283 &4.8 &7.03 &3.5 &6.4 &$\gtsimeq 13$\\

IRTF \ \ &1986 Jul 30.0\ \ \ &581.9\ \ \

&[Ne VI] 7.64 &247$\pm$19 &4.0

&2.720 &4.8 &3.44 &0.58&5.5 &$\gtsimeq 3.6$\\

WIRO \ \ &1986 Aug 11.0\ \ \ &593.0\ \ \

&[Ne VI] 7.64 &249$\pm$25 &4.0

&2.720 &4.8 &3.44 &0.58 &5.6 &$\gtsimeq 3.6$\\

HALE \ \ &1992 Aug 09.3\ \ \ &2783.9\ \ \

&[Ne II] 12.81 &6.84$\pm$1.09 &3.0

&0.272 &0.04 &2.31 &24.1 &0.61 &$\gtsimeq 76$\\

SST &2004 May 11.7\ \ \ &7077.6\ \ \

&[Ne II] 12.81 &0.475$\pm$0.032 \tablenotemark{b} &3.0

&0.272 &0.0026 &2.31 &23.1 &0.04 &$\gtsimeq 73$\\

SST &2004 May 11.7\ \ \ &7077.6\ \ \

&[Ne III] 15.56 &1.09$\pm$0.028 &3.0

&0.596 &0.0026 &1.08 &30.1 &0.17 &$\gtsimeq 95$\\

\hline \\

SST &2004 May 11.7\ \ \ &7077.6\ \ \

&Total Neon&--&--

&--&--&--&--&--&$\gtsimeq 168$\\

\hline \\

SST\ \ \ &2004 May 11.7\ \ \ &7077.6\ \ \

&[O IV] 25.91 &0.304$\pm$0.012 &3.0

&1.641 &0.0026 &0.02 &4.9 &0.001 &$\gtsimeq 2.3$\\

\enddata

\tablenotetext{a}{ Column 12 represents the abundance by number of the given
emission line.}

\tablenotetext{b}{ The IRS 12.81~\micron \ \ values are the weighted mean of
measurements from the SL and SH modules.}

\end{deluxetable}


\clearpage

\clearpage

\clearpage

\clearpage

\clearpage

\end{document}